\documentclass[aps,prb,showpacs,tightenlines,twocolumn,nofootinbib,nobibnotes,superscriptaddress]{revtex4-1}
\usepackage{amsmath,amssymb,amsfonts,bm}
\usepackage{graphicx}
\usepackage{epstopdf}
\usepackage{dcolumn}
\usepackage{mathrsfs}
\usepackage[colorlinks=true,linkcolor=blue,citecolor=blue, urlcolor=blue,bookmarks=false]{hyperref}

\begin{document}

\title{Floquet Weyl semimetals in light-irradiated type-II and hybrid line-node
semimetals}
\date{\today }
\author{Rui Chen}
\author{Bin Zhou}
\thanks{binzhou@hubu.edu.cn}
\author{Dong-Hui Xu}
\thanks{donghuixu@hubu.edu.cn}
\affiliation{Department of Physics, Hubei University, Wuhan 430062, China}
\begin{abstract}
Type-II Weyl semimetals have recently attracted intensive research interest
because they host Lorentz-violating Weyl fermions as
quasiparticles. The discovery of type-II Weyl semimetals evokes the study of
type-II line-node semimetals (LNSMs) whose linear dispersion is strongly
tilted near the nodal ring. We present here a study on the circularly
polarized light-induced Floquet states in type-II LNSMs, as well as those in
hybrid LNSMs that have a partially overtilted linear dispersion in the
vicinity of the nodal ring. We illustrate that two distinct types of Floquet
Weyl semimetal (WSM) states can be induced in periodically driven type-II
and hybrid LNSMs, and the type of Floquet WSMs can be tuned by the direction
and intensity of the incident light. We construct phase diagrams of
light-irradiated type-II and hybrid LNSMs which are quite distinct from
those of light-irradiated type-I LNSMs. Moreover, we show that
photoinduced Floquet type-I and type-II WSMs can be characterized by the
emergence of different anomalous Hall conductivities.
\end{abstract}

\maketitle

\section{Introduction}

Topological semimetals represent a new class of topological matter, which
is characterized by a gapless bulk with a nontrivial band topology. A Weyl
semimetal (WSM) is a kind of topological semimetal that supports Weyl
fermions as low-energy excitations. According to the electronic band structures,
WSMs can be divided into three distinct types: a type-I WSM that has a
pointlike Fermi surface \cite%
{WanX11PRB,YangKY11PRB,Huang15NatCommun,Weng15PRX}, a type-II WSM whose
Fermi surface consists of an electron pocket and a hole pocket touching at
the Weyl nodes \cite{SoluyanovAA15Nature,XuY15PRL}, and a hybrid WSM in
which one Weyl node belongs to type I whereas its chiral partner belongs to
type II \cite{LiFY16PRB}. Earlier research interests were mainly
concentrated on type-I WSMs since real type-I WSM materials had been
theoretically proposed \cite{Huang15NatCommun,Weng15PRX} and experimentally
confirmed in inversion-symmetry-breaking TaAs-class crystals\cite%
{XuSY15Science,LvBQ15PRX,XuSY15NatPhys,Yang15NatPhys}. When Lorentz
invariance is broken, Weyl cones may be tipped over and transformed into
type II. Recently, promising materials such as MoTe$_2$ and WTe$_2$ have
been proposed to be type-II WSMs\cite%
{SoluyanovAA15Nature,WangC16PRB,BrunoFY16PRB,WuY16PRB,FengB16PRB,SunY15PRB,WangZ16PRL}%
, and experimental confirmations of MoTe$_2$ have been reported\cite%
{DengK16NatPhys,Huang16NatMaterials}. Additionally, it was reported that we
can convert TaAs and WTe$_2$ into hybrid WSMs by doping with magnetic ions
and creating magnetic orders in them \cite{LiFY16PRB}.

Another kind of topological semimetal is the so-called line-node semimetal
(LNSM). Unlike WSMs in which the conduction band touches the valence band at
discrete points in momentum space, in LNSMs the conduction band and the
valence band touch along lines. In analogy to WSMs, according to the tilting
degree of the band spectra around the nodal rings, LNSMs can also be classified into
type-I, type-II, and hybrid categories. To date, type-I LNSMs have been
intensively studied both theoretically \cite%
{Heikkila11JETP,Burkov11PRB,Chiu11PRB,ChanYH16PRB,Chiu15PRB,Fang15PRB,Ali14PRB,Xie15APL,Kim15PRL,Yu15PRL,Bian16PRB, Weng16JPCM,Huang16PRB,Mullen15PRL,Bian16NatComm,Schoop16NatCommun}
and experimentally \cite%
{Xie15APL,Bian16NatComm,Schoop16NatCommun,Neupane16PRB,Wu16NatPhys,Hu16PRL,Takane16PRB}%
, however, research on type-II LNSMs is just beginning \cite%
{Hyart16PRB,Volovik17JLTP, LiS17PRB,ZhangX17JPCL,HeJ17Arxiv,GaoY17Arxiv}.
The very recent angle-resolved photoemission spectroscopy measurements on Mg$%
_3$Bi$_2$ suggest it to be a promising candidate material for a type-II LNSM \cite%
{Chang17Arxiv}. As far as we know, the hybrid LNSM, characterized by a
partially overtilted linear dispersion in the vicinity of the nodal ring,
has yet to be proposed and studies are lacking. Type-I LNSMs exhibit intriguing
physical phenomena such as a three-dimensional (3D) quantum Hall effect \cite%
{Mullen15PRL}, 3D flat Landau levels \cite{Rhim15PRB}, $n^{1/4}$
dependence of the plasmon frequency on the charge concentration in the long-wavelength limit \cite{Rhim16NJP,Yan16PRB}, and a quasitopological
electromagnetic response \cite{Ramamurthy17PRB}. Owing to peculiar band
spectra, type-II and hybrid LNSMs are expected to display more intriguing
phenomena.

Application of light offers a powerful method to manipulate electronic
states, and even change the band topology in solids \cite%
{WangY13Science,Mahmood16NatPhys,SieEJ15NatureMater,KimJ14Science}. A
typical example is the Floquet topological insulator \cite%
{Lindner11NatPhys,Titum17PRB,Titum15PRL}, which is a direct consequence of
changing the band topology by means of light. Moreover, photoinduced
topological states in other two-dimensional systems, such as graphene \cite%
{Oka09PRB,KitagawaT11PRB,GuZ11PRL,Piskunow14PRB} and silicene\cite%
{Ezawa13PRL}, have been studied. Recently, light-driven semimetals have
attracted much attention\cite%
{WangR14EPL,Chan16PRL,Ebihara16PRB,Chan16PRB,YanZ16PRL,Narayan16PRB,Taguchi16PRB1,
Ezawa17PRB,YanZ17PRB1,YaoS17PRB,YanZ17PRB2,Klinovaja17PRL,Bomantara16PRBE,Chan17PRB}%
. It was found that a Floquet WSM phase can be generated from a light-driven
Dirac semimetal due to time-reversal symmetry breaking \cite%
{WangR14EPL,Chan16PRL,Ebihara16PRB,Chan16PRB}. Later, it was shown that
circularly polarized light can drive a type-I LNSM into a WSM, accompanied
by photovoltaic anomalous Hall conductivity \cite%
{Chan16PRB,YanZ16PRL,Narayan16PRB,Taguchi16PRB1,Ezawa17PRB}. A Floquet WSM
phase with multi-Weyl points was also proposed in crossing LNSMs\cite%
{YanZ17PRB1,YaoS17PRB,YanZ17PRB2}.

\begin{figure}[ptb]
\includegraphics[width=8cm]{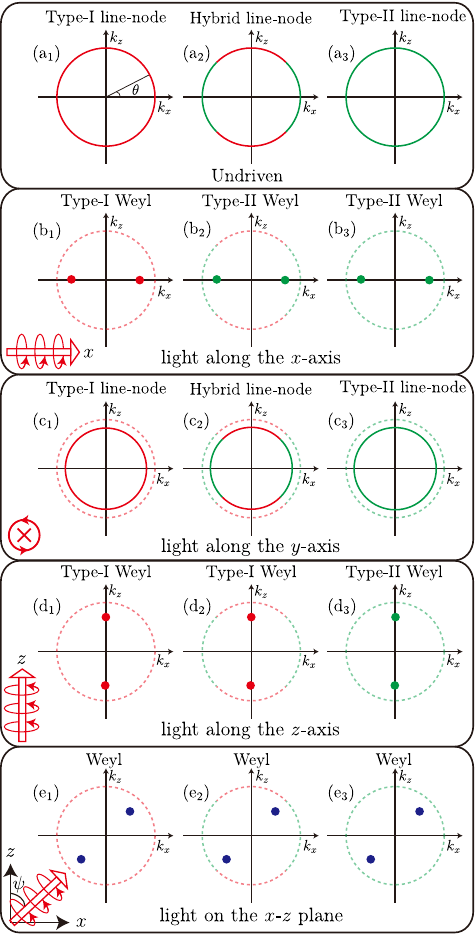}
\caption{(Color online) Schematics of driven type-I, hybrid, and type-II
LNSMs. (a) Nodal rings of undriven LNSMs. (b) and (d) correspond to the case
where the incident light travels along the $x$ axis and $z$ axis,
respectively. The dots colored with red (or green) denote type-I (or II)
Weyl nodes. Nodal rings are gapped out and Weyl nodes are created along the
propagation direction of light. (c) The incident light along the $y$ axis
only shifts the nodal rings. (e) The light propagates on the $x$-$z$ plane, with $%
\protect\psi$ defining the incident angle away from the $z$ axis. Nodal
rings are gapped out and pairs of Weyl nodes appear along the propagation
direction. We label the Weyl nodes with blue color as their type depends on the
model parameters, the incident angle and the strength of light. The dashed
lines in (b)-(e) correspond to nodal rings in the undriven cases.}
\label{fig1}
\end{figure}

In this paper, we present a systematic study on Floquet states in
periodically driven type-II [Fig.~\ref{fig1}(a$_3$)] and hybrid [Fig.~\ref%
{fig1}(a$_2$)] LNSMs by means of light. We show that Floquet WSMs can be
created by applying circularly polarized light. When the incident light
propagates along the $x$ axis or along the $z$ axis, a type-II LNSM is
converted into a type-II WSM [Figs.~\ref{fig1}(b$_3$) and \ref{fig1}(d$_3$%
)], while for a driven hybrid LNSM, depending on the tilt direction, the
photoinduced Floquet WSM could be of type I [Fig.~\ref{fig1}(b$_2$)] or
type II [Fig.~\ref{fig1}(d$_2$)]. When the applied light propagates along
the $y$ axis, only the positions of the nodal rings change [Figs.~\ref{fig1}(c$_2
$) and \ref{fig1}(c$_3$)]. Surprisingly, by rotating incident light on
the $x$-$z$ plane, both type-I and type-II WSMs can be realized by tuning
the driving angle and amplitude [Figs.~\ref{fig1}(e$_2$) and \ref{fig1}(e$_3$%
)]. For the sake of comparison, we also give the Floquet states of driven
type-I LNSMs by circularly polarized light [Figs.~\ref{fig1}(a$_1$)-\ref%
{fig1}(e$_1$)] which show different features from those of type-II and
hybrid LNSMs. We summarize all the results in three distinct phase diagrams
in ($A_L$, $\psi$) space, where $A_L$ and $\psi$ are the amplitude and
the incident angle, respectively. Lastly, by use of the Kubo formula, the
anomalous Hall effect of photoinduced Floquet WSMs is also investigated.

\section{Model}

To study periodically driven LNSMs, we start with a simple two band model of
LNSMs with a single nodal ring. The model Hamiltonian of undriven LNSMs is
written as\cite{Kim15PRL,Yu15PRL,WengH15PRB,Chan16PRB}
\begin{equation}
H_{0}=c_{i}k_{i}^{2}\sigma_{0}+\left( m_{0}-m_{i}k_{i}^{2}\right) \sigma
_{z}+v_{y}k_{y}\sigma_{y},  \label{H0}
\end{equation}
where $m_0, m_i (i=x,y,z)$ and $c_i$ are model parameters, $v_y$ is the
velocity along the $y$-axis, $k_i$ are the crystal momenta, $\sigma_i$ are
Pauli matrices and $\sigma_0$ is the identity matrix. We use Einstein's
summation convention that repeated indices indicate the summation is
implied. This model respects both the time-reversal
and inversion symmetries, thus can be applied to spinless LNSM systems\cite%
{Kim15PRL,WengH15PRB,Yu15PRL}. The eigenvalues are obtained by diagonalizing
the Hamiltonian (\ref{H0}),
\begin{equation}
E_{\pm}^{0}\left( \mathbf{k}\right) =c_{i}k_{i}^{2}\pm\sqrt{\left(
m_{0}-m_{i}k_{i}^{2}\right) ^{2}+v_{y}^{2}k_{y}^{2}}.
\end{equation}
Under the band inversion condition $m_{0,x,y,z}>0$, a nodal ring appears
along an ellipse defined by $m_{x}k_{x}^{2}+m_{z}k_{z}^{2}=m_{0}$ at $k_{y}=0
$. The nodal ring is located at $\mathbf{k}_{0}=\left( \sqrt{m_{0}/m_{x}}%
C_{\theta},0,\sqrt{m_{0}/m_{z}}S_{\theta}\right) , $ where $%
C_{\theta}=\cos\theta$, $S_{\theta}=\sin\theta$, and $\theta$ is the polar
angle shown in Fig.~\ref{fig1}(a$_1$). Linearizing the eigenvalues around
the nodal ring we get the energy dispersion%
\begin{align}
E_{\pm}^{0}\left( \mathbf{q}=\mathbf{k}-\mathbf{k}_{0}\right)& =\boldsymbol{w%
}\cdot\mathbf{q}\pm \sqrt{\left(\xi_{x}q_{x}+\xi_{z}q_{z}\right)^2+%
\xi_{y}^{2}q_{y}^{2}}  \notag \\
& =T\left( \mathbf{q}\right) \pm U\left( \mathbf{q}\right),  \label{E0}
\end{align}
where $\xi_{x}^{2}=4m_{0}m_{x}C_{\theta}^{2}$, $\xi_{y}^{2}=v_{y}^{2}$, $%
\xi_{z}^{2}=4m_{0}m_{z}S_{\theta}^{2}$, and $\boldsymbol{w}=2\left(
c_{x}C_{\theta}\sqrt{m_{0}/m_{x}},0,c_{z}S_{\theta}\sqrt{m_{0}/m_{z}} \right)
$. The first $T$ term of Eq. (\ref{E0}) describes the
tilt of the energy dispersion, and the second $U$ term of Eq. (\ref{E0})
denotes the splitting of the energy band. Around point $\mathbf{k}_0$ on the nodal ring, we obtain a two dimensional Dirac cone when the small vector $\mathbf{q}$ is confined on the transverse plane formed by the inplane normal line to the nodal ring and the line along the $k_y$ axis. From Eq.~\eqref{E0}, we can see that the tilt is most effective when $\mathbf{q}$ is along the direction of normal vector $\mathbf{k}_n=(m_xk_{0x},m_zk_{0z})$,
then the tilt radio can be defined by \cite%
{SoluyanovAA15Nature,LiS17PRB}
\begin{align}  \label{tilt2}
F_{\theta}  =\left\vert \frac{T( \hat{\mathbf{k}}_{n}) }{U( \hat{\mathbf{k}}_{n}) }\right\vert
=\left\vert\frac{c_{x}C_{\theta}^{2}+c_{z}S_{\theta}^{2}}{m_{x}C_{\theta}^{2}+m_{z}S_{\theta}^{2}}\right\vert,
\end{align}
where $\hat{\mathbf{k}}_n=\mathbf{k}_n/|\mathbf{k}_n|$ is the normal unit vector on the $k_x-k_z$ plane. The nodal ring is of type-I when $F_{\theta}<1$ for all the values of $\theta
$, and of type-II when $F_{\theta}>1$ for all the values of $\theta$. At the
intersection points between the nodal ring and the $x$-axis, the tilt ratios
are $F_{0,\pi}=\left\vert c_{x}/m_{x}\right\vert $. For the intersection
points between the nodal ring and the $z$-axis, the tilt ratios become $%
F_{\pi/2,3\pi/2}=\left\vert c_{z}/m_{z}\right\vert $. When $\left\vert
c_{x}/m_{x}\right\vert >1$, $\left\vert c_{z}/m_{z}\right\vert >1$, and $c_xc_z>0$, it is
easy to find that $F_{\theta}$ is always greater than $1$, then the system
is a type-II LNSM [Fig. \ref{fig1}(a$_3$)]. While for $\left\vert
c_{x}/m_{x}\right\vert <1$ and $\left\vert c_{z}/m_{z}\right\vert <1$, $%
F_{\theta}$ is always smaller than $1$, then the system is a type-I LNSM
[Fig.~\ref{fig1}(a$_1$)]. For the rest cases, depending on $\theta$, $%
F_{\theta}$ on the nodal ring may be greater or smaller than 1, we call it a
hybrid LNSM [Fig.~\ref{fig1}(a$_2$)].

\begin{figure}[ptb]
\includegraphics[width=8cm]{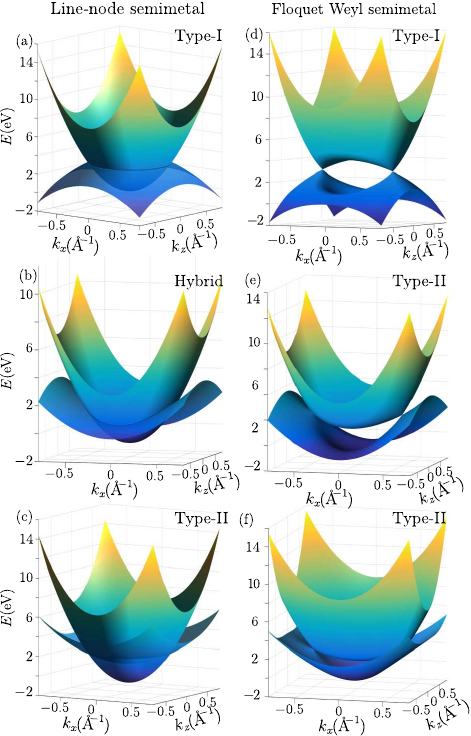}
\caption{(Color online) (a)-(c) are the energy spectra of undriven
type-I, hybrid, and type-II LNSMs with (a) $\left\vert c_{x}/m_{x}\right\vert
=\left\vert c_{z}/m_{z}\right\vert <1$, (b) $\left\vert
c_{x}/m_{x}\right\vert >1$ and $\left\vert c_{z}/m_{z}\right\vert <1$, and
(c) $\left\vert c_{x}/m_{x}\right\vert =\left\vert c_{z}/m_{z}\right\vert >1$%
, where we take $k_y=0$. (d)-(f) are the energy spectra of Floquet WSMs in
light-driven type-I, type-II, and hybrid LNSMs along the $x$ axis.}
\label{fig2}
\end{figure}

The above analytical results can be illustrated more clearly by plotting the
bulk spectra in Figs.~\ref{fig2}(a)-\ref{fig2}(c). As shown in Fig.~\ref%
{fig2}(a), the tilt is weak, the band touching forms a type-I nodal ring. In
Fig.~\ref{fig2}(c), the tilt is strong enough such that both bands radiate
the same direction, and their intersection makes a type-II nodal ring.
Figure~\ref{fig2}(b) shows band spectrum for a hybrid LNSM, we can see that
the tilt ratio $F_{\theta}$ is smaller than 1 near the $k_z$-axis, and the
ratio $F_{\theta}$ is greater than 1 near the $k_x$-axis.

\section{Floquet states}

To study the interaction of LNSMs with light, we consider a time-dependent
vector potential $\mathbf{A}\left( t\right) =\mathbf{A}\left( t+T\right) $,
which is a periodic function with a period of $T=2\pi/\omega$. The driven
Hamiltonian $H\left( t\right) $ is obtained by using the Peierls
substitution, $\mathbf{k}\rightarrow \mathbf{k}-\mathbf{A}(t)$. In this paper,
we focus on the low-energy physics of LNSMs near the nodal ring.
Making use of Floquet theory\cite{Shirley65PRB,Sambe73PRA,Gesztesy81JPMG}
in the high-frequency limit, the periodically driven system can be
described by a static effective Hamiltonian as\cite%
{Maricq82PRB,Grozdanov88PRB,Rahav03PRA,Rahav03PRL,Goldman14PRX,Eckardt15NJP,Bukov15AdvPhys}
\begin{equation}
H_{\text{eff}}=H_{0,0}+\frac{\left[ H_{0,-1},H_{0,1}\right] }{\hbar \omega}%
+O\left( A_{L}^{4}\right) ,  \label{effective}
\end{equation}
where $\omega$ and $A_L$ describe the frequency and amplitude of
light, and $H_{m,n}=\frac{1}{T}\int_{0}^{T}H (t) e^{i(m-n)\omega t}dt $ are the
discrete Fourier components of the Hamiltonian.

\subsection{Light propagating along the $x$ axis}

\label{lightx} When a light propagates along the $x$ axis, $\mathbf{A}$ is
given by $\mathbf{A}=A_{L}\left( 0,\cos\omega t,\eta\sin\omega t\right)$,
where $\eta=\pm1$ indicates the chiralities of the circularly polarized
light. From Eq. (\ref{effective}), the Floquet correction is
\begin{equation}
\Delta H^{x}=-\frac{A_{L}^{2} }{2}\left(
m_{y}+m_{z}\right)\sigma_{z}-Lm_{z}k_{z}\sigma_{x},
\end{equation}
with $L=2\eta A_{L}^{2}v_{y}/(\hbar\omega)$. We obtain a term coupling the momentum
$k_z$ and $\sigma_x$, which gaps out the nodal ring except at two Weyl
points $\pm\mathbf{k}_{0}=( \pm\sqrt{\tilde{m}_{0}/m_{z}},0,0)$ with $\tilde{%
m}_{0}=m_{0}-A_{L}^{2}\left( m_{y}+m_{z}\right)/2$.  Linearizing the
eigenvalues of Hamiltonian around $\mathbf{k}_{0}$, we have the energy
dispersion%
\begin{align}
E_{\pm}\left( \mathbf{q}\right) &=\frac{2c_{x}\sqrt{\tilde{m}_{0}}}{\sqrt{%
m_{x}}}q_{x}\pm\sqrt{4\tilde{m}_{0}m_{x}q_{x}^{2}+\left( Lm_{z}q_{z}\right)
^{2}+v_{y}^{2}q_{y}^{2}}  \notag \\
&= T(\mathbf{q})\pm U(\mathbf{q}),
\end{align}
where $\mathbf{q}=\mathbf{k}-\mathbf{k}_{0}$. The energy dispersion near $-%
\mathbf{k}_0$ is $E_{\pm}\left( \mathbf{q}\right)=-T(\mathbf{q})\pm U(%
\mathbf{q})$. The band is tilted along the $x$ axis, and then the tilt ratio of
the Weyl points is given by
\begin{equation}
F_x=\left\vert \frac{T\left( q_{x}\right) }{U\left( q_{x}\right) }%
\right\vert =\frac{c_{x}}{m_{x}}.
\end{equation}
Interestingly, the type of the pair of Weyl nodes only depends on the ratio $%
c_x/m_x$ and has nothing to do with the intensity and the frequency of the
applied light.

The results above show that when light traveling along the $x$ axis gaps out
the nodal ring and leaves a pair of Weyl nodes, the system enters into a WSM
phase. However, the type of the Weyl nodes is independent of the intensity
and the frequency of the incident light, which implies that a type-II
Floquet WSM state arises by driving the type-II LNSM with light along the $%
x$ axis [Fig.~\ref{fig1}(b$_3$)] since $c_x/m_x>1$ for the type-II LNSM. For
the driven hybrid LNSM, the type of induced Weyl nodes depends on the
specific value of $c_x/m_x$ [Fig.~\ref{fig1}(b$_2$)], that is to say, a
type-I WSM state arises if $c_x/m_x<1$ and a type-II WSM state appears if $%
c_x/m_x>1$.

The bulk band spectra of the driven hybrid LNSM with $c_x/m_x>1$ and the
driven type-II LNSM are shown in Figs.~\ref{fig2}(e) and \ref{fig2}(f),
respectively. It can be seen that the type-II Weyl nodes are separated along
the propagation direction of the incident light, as predicted. For
comparison, we also plot the band spectrum of the driven type-I LNSM [Fig.~%
\ref{fig2}(d)] showing the type-I Weyl nodes, which were revealed in previous
studies\cite{YanZ16PRL,Narayan16PRB,Taguchi16PRB1,Ezawa17PRB}.

\subsection{Light propagating along the $y$ axis}

When the incident light propagates along the $y$ axis, $\mathbf{A}$ is given
by $\mathbf{A}=A_{L}\left( \eta\sin\omega t,0,\cos\omega t\right)$, and it
produces the following correction,
\begin{equation}
\Delta H^{y}=-\frac{A_{L}^{2} }{2}\left( m_{x}+m_{z}\right)\sigma_{z}.
\end{equation}
The correction term can be absorbed in the second term of Eq.~(\ref{H0}) by
renormalizing the parameter $m_{0}\rightarrow m_{0}-A_{L}^{2}\left(
m_{x}+m_{z}\right)/2$. It means that the incident light propagating along
the $y$ axis only shifts the nodal rings instead of gapping them out [Figs. %
\ref{fig1}(c$_1$)-\ref{fig1}(c$_3$)].

\subsection{Light propagating along the $z$ axis}

\label{lightz} For light propagating along the $z$ axis, we have $\mathbf{A%
}=A_{L}\left( \cos\omega t,\eta\sin\omega t,0\right)$. Then the effective
Hamiltonian gains additional terms,
\begin{equation}
\Delta H^{z}=-\frac{A_{L}^{2} }{2}\left(
m_{x}+m_{y}\right)\sigma_{z}+Lm_{x}k_{x}\sigma_{x}.
\end{equation}
In this case, the light gaps out the nodal ring except at two Weyl points $%
\pm\mathbf{k}_{0}=(0,0,\pm\sqrt{\tilde{m}_{0}^{\prime}/m_{z}})$ with $\tilde{%
m}_{0}^{\prime}=m_{0}-A_{L}^{2}\left( m_{x}+m_{y}\right)/2$, which is
similar to the case of light propagating along the $x$ axis. Linearizing the
eigenvalues of Hamiltonian around $\mathbf{k}_{0}$, we have the energy
dispersion%
\begin{align}
E_{\pm}\left( \mathbf{q}\right) &=\frac{2c_{z}\sqrt{\tilde{m}_{0}^{\prime}}}{%
\sqrt{m_{z}}}q_{z}\pm\sqrt{4\tilde{m}_{0}^{\prime}m_{z}q_{z}^{2}+\left(
Lm_{x}q_{x}\right) ^{2}+v_{y}^{2}q_{y}^{2}}  \notag \\
&= T(\mathbf{q})\pm U(\mathbf{q}).
\end{align}
Thus the tilt ratio is
\begin{equation}
F_z=\left\vert \frac{T\left( q_{z}\right) }{U\left( q_{z}\right) }%
\right\vert =\frac{c_{z}}{m_{z}}.
\end{equation}
We can conclude that, in the presence of light propagating along the $z$ axis, a LNSM evolves into a WSM with a pair of Weyl nodes separated along
the propagating direction. The type of Weyl nodes is only determined by
the ratio $c_z/m_z$ [Figs.~\ref{fig1}(d$_1$)-\ref{fig1}(d$_3$)].

\subsection{Light propagating on the $x$-$z$ plane}

\begin{figure*}[ptb]
	\includegraphics[width=15cm]{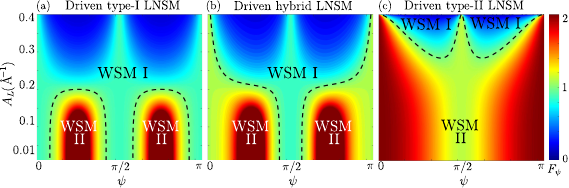}
	\caption{(Color online) Phase diagrams of driven LNSMs in the $(A_L,\protect%
		\psi)$ phase space. The color bar indicates the tilt ratio $F_\protect\psi$.
		The phase boundaries are marked by the dashed lines, corresponding to $F_%
		\protect\psi=1$. The model parameters are taken to be (a) $-c_x=c_y=c_z=3.5 \text{ eV \r{A}}^2$, (b) $%
		-c_x=c_y=3.5 \text{ eV \r{A}}^2$ and $c_z=4.5 \text{ eV \r{A}}^2$, (c) $c_x=c_y=4.5\text{ eV \r{A}}^2$ and $c_z=8\text{ eV \r{A}}^2$.
		The common parameters are $m_x=m_y=m_z=4\text{ eV \r{A}}^2$, $m_0=1\text{ eV} $, $v_y=1\text{ eV \r{A}}$,
		and $\hbar\omega=0.1\text{ eV}$.}\label{fig3}
\end{figure*}

We rotate the propagation direction of the incident light on the $x$-$z$
plane with $\mathbf{A}=A_{L}\left( C_{\psi}\cos\omega t,\eta\sin\omega
t,-S_{\psi}\cos\omega t\right)$, where $C_\psi=\cos\psi$, $S_\psi=\sin\psi$,
and $\psi$ defines the incident angle off the $z$ axis. When $\psi=0,$ the
propagation direction is along the $z$ axis, which is just the case in Sec.~%
\ref{lightz}. When $\psi=\pi/2,$ the incident direction is along the $x$%
-axis, which is the case discussed in Sec.~\ref{lightx}. For generic $\psi$,
the light induces the following Floquet correction,
\begin{align}
\Delta H^{xz} & =-\frac{A_{L}^{2}}{2}\left( C_{\psi}^{2}m_{x}+m_{y}+S_{\psi
}^{2}m_{z}\right) \sigma_{z}  \notag \\
& +L\left( C_{\psi}m_{x}k_{x}-S_{\psi}m_{z}k_{z}\right) \sigma _{x}.
\label{Hxz}
\end{align}
The second term of Eq.~(\ref{Hxz}) is proportional to $\sigma_x$, giving
rise to a pair of Weyl nodes at
\begin{equation}
\pm\mathbf{k}_{0}=\pm\left( \sqrt{\frac{m_{z}}{m_{x}}}S_{\psi},0,\sqrt{\frac{%
m_{x}}{m_{z}}}C_{\psi}\right) \sqrt{\frac{\tilde{m}_{0}^{\prime\prime}}{%
C_{\psi}^{2}m_{x}+S_{\psi}^{2}m_{z}}},
\end{equation}
where $\tilde{m}_{0}^{\prime\prime}=m_{0}-A_{L}^{2}\left(
C_{\psi}^{2}m_{x}+m_{y}+S_{\psi}^{2}m_{z}\right)/2 $. Using the same
procedure in Secs.~\ref{lightx} and \ref{lightz}, the tilt ratio of the Weyl
nodes can be expressed as
\begin{equation}
F_{\psi }=\frac{\lambda _{x}^{2}+\lambda _{z}^{2}}{\sqrt{%
L^{2}P_{1}^{2}+4m_{x}m_{z}\tilde{m}_{0}^{\prime \prime }P_{2}^{2}/K}},
\end{equation}
where $P_{1}=C_{\psi }m_{x}\lambda _{x}-S_{\psi }m_{z}\lambda _{z}$, $%
P_{2}=S_{\psi }\lambda _{x}+C_{\psi }\lambda _{z}$, and $K=m_xC_{\psi}^{2}+m_zS_{\psi}^{2}$. We can see
that, in this case, the tilt ratio depends on the intensity, frequency, and
incident angle of the light, which is quite different from previous cases in
which the type of induced Weyl nodes is only determined by the
parameters of the original Hamiltonian of LNSMs. This allows us to control
the type of Floquet WSMs by tuning the intensity and incident angle of
the applied light [Figs.~\ref{fig1}(e$_1$)-\ref{fig1}(e$_3$)]. It implies
that we can have a type-I Floquet or a type-II WSM by driving a type-II LNSM
with an appropriate light.

Figures \ref{fig3}(a)-\ref{fig3}(c) show the phase diagrams in ($A_L$, $%
\psi$) space for driven type-I, hybrid, and type-II LNSMs obtained by
monitoring the tilt ratio of the Weyl nodes. The phase diagrams show
peculiar features for distinct types of LNSMs. When the light intensity is
weak, the driven type-I LNSM [Fig.~\ref{fig3}(a)] supports the type-I WSM
phase near $\psi=0$, $\pi/2$, and $\pi$, and the type-II WSM phase near $%
\psi=\pi/4$ and $3\pi/4$. However, for the driven hybrid LNSM, the type-I
WSM phase only occupies a small region near $\psi=\pi/2$ in the phase
diagram, and the rest of the phase diagram is occupied by the type-II WSM
phase [Fig.~\ref{fig3}(b)]. This is because we choose a hybrid LNSM with $%
c_z/m_z>1$ and $c_x/m_x<1$.  In contrast to the driven type-I and hybrid
LNSMs, the driven type-II LNSM hosts only the type-II WSM phase at weak
light intensity [Fig.~\ref{fig3}(c)]. As the intensity increases, the type-I
WSM phase dominates the phase diagrams for all types of driven LNSMs.

Now, we discuss the possibility of an experimental realization of Floquet WSMs
in periodically driven LNSMs. Let us consider the realistic parameters $v_{y}=0.6\text{ eV \r{A}}$,
$m_{x,y,z}=10\text{ eV \r{A}}^2$, and $m_{0}\approx 0.3$ eV for a candidate type-II LNSM material
K$_4$P$_3$ \cite{LiS17PRB}. We choose the photon energy to be
$\hbar \omega\approx$ 150 meV, which is close to the typical values in recent optical
pump-probe experiments. The amplitude of light $A_L$ for the onset of a type-I Floquet WSM is
about $0.05\text{ \r{A}}^{-1}$, and the corresponding electric field strength
$E_0= \hbar \omega A_L/e$ is $5\times 10^7$ V/m,
which is within experimental accessibility \cite{WangY13Science,Mahmood16NatPhys},
while a type-II WSM phase in the driven system can appear even at very weak light intensity.

\section{Photoinduced anomalous hall effect}

The photoinduced phase transition from LNSMs to WSMs is accompanied by an
anomalous Hall effect since time-reversal breaking WSMs exhibit nonzero,
nonquantized Hall conductivity \cite{YangKY11PRB,Burkov14PRL}. In this
section, we study the photovoltaic anomalous Hall effect of Floquet WSM
states in driven LNSMs. We will concentrate on the case in which the
incident light propagates along the $x$ axis. The Weyl nodes are located
along the $x$-axis, thus the nontrivial component of Hall conductivity is $%
\sigma_{zy}^{\text{AHE}}$, which can be obtained by use of the Kubo formula%
\cite{Taguchi16PRB1,Oka09PRB}
\begin{align}
\sigma_{zy}^{\text{AHE}} & =-i\hbar e^2\int\frac{d^{3}\mathbf{k}}{\left(
2\pi\right) ^{3}}\sum_{\alpha\neq\beta}\frac{f[\epsilon_{F}-E_{\beta}\left(
\mathbf{k}\right) ]-f[\epsilon_{F}-E_{\alpha}\left( \mathbf{k}\right) ]}{%
E_{\beta}\left( \mathbf{k}\right) -E_{\alpha}\left( \mathbf{k}\right) }
\notag \\
& \times\frac{\left\langle \psi_{\alpha}\left( \mathbf{k}\right) \left\vert
v_{z}\left( \mathbf{k}\right) \right\vert \psi_{\beta}\left( \mathbf{k}%
\right) \right\rangle \left\langle \psi_{\beta}\left( \mathbf{k}\right)
\left\vert v_{y}\left( \mathbf{k}\right) \right\vert \psi_{\alpha}\left(
\mathbf{k}\right) \right\rangle }{E_{\beta}\left( \mathbf{k}\right)
-E_{\alpha}\left( \mathbf{k}\right) +i\delta},  \label{hall}
\end{align}
where $f$ is the Fermi distribution function, $v_{z,y}\left( \mathbf{k}%
\right) =\left[ \partial H\left( \mathbf{k}\right) /\partial k_{z,y}\right]%
/\hbar$ are the velocity operators along the $z$  and $y$ axes, $\delta$ is
an infinitesimal quantity, $E_{\alpha}\left( \mathbf{k}\right)$ is the $%
\alpha$th band of the effective Hamiltonian, and $\psi_{\alpha}(\mathbf{k})$
is the corresponding eigenvector. The anomalous Hall conductivity is easily
obtained when the Fermi energy is located at the Weyl nodes ($\epsilon_{F}=0$%
) and the temperature is zero. For ideal type-I WSMs, the anomalous Hall
conductivity is $\sigma_{\text{type-I}}^{\text{AHE}}=e^2 Q/(h\pi)$\cite%
{Burkov11PRL,Burkov11PRB}, where $2Q$ is the distance between the Weyl nodes
in momentum space. For type-II WSMs, due to the strong tilt of the Weyl nodes,
the anomalous Hall conductivity is found to be related to the tilt ratio
\cite{Zyuzin16JETPL}.

\begin{figure}[ptb]
\includegraphics[width=8cm]{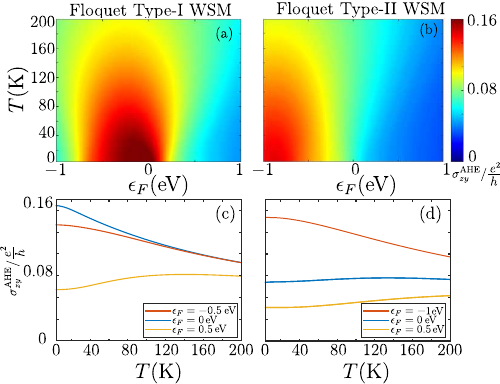}
\caption{(Color online) The anomalous Hall conductivity $\protect\sigma%
_{zy}^{\text{AHE}}$ as functions of the Fermi energy $\protect\epsilon_{F}$
and temperature $T$ for (a) the Floquet type-I WSM and (b) the Floquet
type-II WSM. The anomalous Hall conductivities $\protect\sigma_{zy}^{\text{%
AHE}}$ as a function of the temperature $T$ at different Fermi energies $%
\protect\epsilon_{F}$ for (c) the type-I WSM and (d) the type-II WSM. The
common parameters are $m_x=m_y=m_z=4\text{ eV \r{A}}^2$, $m_0=1\text{ eV}$, $v_y=1\text{ eV \r{A}}$, $\hbar\omega=0.1
\text{ eV}$, and $A_L=0.3\text{ \r{A}}^{-1}$. The parameters for the Floquet type-I WSM are $%
c_x=c_y=c_z=2\text{ eV \r{A}}^2$, and for the Floquet type-II WSM are $c_x=c_y=c_z=8\text{ eV \r{A}}^2$.}
\label{fig4}
\end{figure}

For arbitrary Fermi energy $\epsilon_{F}$ and temperature $T$, the anomalous
Hall conductivity can be numerically calculated by using Eq.~(\ref{hall}).
As shown in Fig.~\ref{fig4}, we calculate the anomalous Hall conductivity in photoinduced Floquet type-I and type-II WSMs, which correspond to the
case shown in Figs.~\ref{fig2}(d) and \ref{fig2}(f), respectively. For the
Floquet type-I WSM at low temperatures [Fig.~\ref{fig4}(a)], the anomalous
Hall conductivity reaches its maximum near $\epsilon_{F} = 0$, and reduces
at finite Fermi energies, whereas for the Floquet type-II WSM at low
temperatures [Fig.~\ref{fig4}(b)], the maximum value of the anomalous Hall
conductivity occurs at a Fermi energy $\epsilon_{F}=-1\text{ eV}$ due to the
imbalance between the electron and hole pockets. Note that, the Hall
conductivity of the type-I Floquet WSM is asymmetric with respect to the
Fermi energy, which is attributed to the weak tilt of the energy dispersion.
Because of the anisotropy of the band structure caused by the strong tilt in the
energy dispersion, $\sigma_{zy}^{\text{AHE}}$ in the Floquet type-II WSM is
highly asymmetric with respect to the Fermi energy.

We plot the anomalous Hall conductivities of the Floquet type-I and type-II
WSMs as a function of the temperature $T$ with different Fermi energies in
Figs.~\ref{fig4}(c) and \ref{fig4}(d). When the Fermi energy is located at
the Weyl nodes, i.e., $\epsilon_{F} = 0$, the Hall conductivity $%
\sigma_{zy}^{\text{AHE}}$ for the Floquet type-I WSM decreases with
increasing temperature, however, it increases as temperature increases for
the Floquet type-II WSM. When the Fermi energy is located below the energy
of Weyl nodes, $\epsilon_{F}=-0.5\text{ eV}$ for the type-I Floquet WSM and $%
\epsilon_{F}=-1\text{ eV}$ for the type-II Floquet WSM, $\sigma_{zy}^{\text{AHE}}$
decreases as the temperature increases. When the Fermi energy is located
above the energy of Weyl nodes $\epsilon_{F}=0.5\text{ eV}$, $\sigma_{zy}^{\text{AHE}}$
increases as the temperature increases for both type-I and type-II
Floquet WSMs. It is necessary to point out that the results are obtained in
the low-temperature regime, and we ignore the high-temperature case where $%
\sigma_{zy}^{\text{AHE}}$ finally decreases to zero.

\section{Conclusion}

In this paper, we identify a different type of LNSM, a hybrid LNSM, and investigate
the effect of off-resonant circularly polarized light on type-II and hybrid LNSMs
within the framework of Floquet theory. We show that both of them
can support photoinduced Floquet WSM phases. Remarkably, we can manipulate distinct types
of WSM states by tuning the incident angle and amplitude of light.
Type-II and hybrid LNSMs, along with type-I LNSMs,
provide highly controllable platforms for creating WSM states.
We also study the anomalous Hall effect of driven LNSMs, which can be used to
characterize different types of photoinduced LNSM-WSM transitions.

In comparison with other proposals for realizing artificial WSM phases, such as a magnetically
doped topological insulator multilayer \cite{Burkov11PRL}, driving LNSMs with circularly polarized light is
a promising alternative way to realize distinct types of WSM phases without fine tuning,
and it does not introduce disorder. The Floquet WSM states in periodically driven LNSMs are ready to be realized,
considering the Floquet-Bloch states have been successfully observed on the surface of the topological insulator Bi$_2$Se$_3$
by the use of time- and angle-resolved photoemission spectroscopy \cite{Mahmood16NatPhys,WangY13Science}.
The anomalous Hall effect associated with Floquet WSMs can be detected by transport measurements.

\section*{Acknowledgments}

R.C. and D.-H.X. were supported by the National Natural Science Foundation
of China (Grant No. 11704106). D.-H.X. also acknowledges the support of
Chutian Scholars Program in Hubei Province. B.Z. was supported by the
National Natural Science Foundation of China (Grant No. 11274102), the
Program for New Century Excellent Talents in University of Ministry of
Education of China (Grant No. NCET-11-0960), and the Specialized Research
Fund for the Doctoral Program of Higher Education of China (Grant No.
20134208110001).

\end{document}